\documentclass[sigconf]{acmart}

\AtBeginDocument{%
  }

\setcopyright{acmlicensed}
\copyrightyear{2018}
\acmYear{2018}
\acmDOI{XXXXXXX.XXXXXXX}

\acmConference[Conference acronym 'XX]{Make sure to enter the correct
  conference title from your rights confirmation emai}{June 03--05,
  2018}{Woodstock, NY}
\acmISBN{978-1-4503-XXXX-X/18/06}

\begin{document}

\title{Designing VR Simulation System for Clinical Communication Training with LLMs-Based Embodied Conversational Agents}


\author{Xiuqi Tommy Zhu}
\email{zhu.xiu@northeastern.edu}
\affiliation{
  \institution{Northeastern University}
  \country{United States}
}

\author{Heidi Cheerman}
\email{h.cheerman@northeastern.edu}
\affiliation{
  \institution{Northeastern University}
  \city{Boston}
  \country{United States}
}

\author{Minxin Cheng}
\email{cheng.min@northeastern.edu}
\affiliation{
  \institution{Northeastern University}
  \city{Boston}
  \country{United States}
}
\author{Sheri Kiami}
\email{s.kiami@northeastern.edu}
\affiliation{
  \institution{Northeastern University}
  \city{Boston}
  \country{United States}
}
\author{Leanne Chukoskie}
\email{l.chukoskie@northeastern.edu}
\affiliation{
  \institution{Northeastern University}
  \city{Boston}
  \country{United States}
}
\author{Eileen McGivney}
\email{e.mcgivney@northeastern.edu}
\authornote{Corresponding Author}
\affiliation{
  \institution{Northeastern University}
  \city{Boston}
  \country{United States}
}

\renewcommand{\shortauthors}{Zhu et al.}
\renewcommand{\shorttitle}{Designing VR Simulation System for Clinical Communication Training with
LLMs-Based Embodied Conversational Agents}
\begin{abstract}
  VR simulation in Health Professions (HP) education demonstrates huge potential, but fixed learning content with little customization limits its applicationbeyond lab environments. To address these limitations in the context of VR for patient communication training, we conducted a user-centered study involving semi-structured interviews with advanced HP students to understand their challenges in clinical communication training and perceptions of VR-based solutions. From this, we derived design insights emphasizing the importance of realistic scenarios, simple interactions, and unpredictable dialogues. Building on these insights, we developed the Virtual AI Patient Simulator (VAPS), a novel VR system powered by Large Language Models (LLMs) and Embodied Conversational Agents (ECAs), supporting dynamic and customizable patient interactions for immersive learning. We also provided an example of how clinical professors could use user-friendly design forms to create personalized scenarios that align with course objectives in VAPS and discuss future implications of integrating AI-driven technologies into VR education. 

\end{abstract}

\ccsdesc[300]{Human-centered computing~Virtual Reality}
\ccsdesc[500]{Applied computing~Interactive learning environments}

\keywords{Virtual Reality, Embodied Conversational Agents, Clinical Simulation, Medical Education}

\received{20 February 2007}
\received[revised]{12 March 2009}
\received[accepted]{5 June 2009}

\maketitle

\section{Introduction}
Healthcare professionals, including nurses, physical therapists, physician's assistants, and pharmacists, are increasingly in demand as the United States population grows and ages \cite{Workforce2024}, making health profession (HP) education crucial. These professions require frequent direct interaction with patients with diverse needs, backgrounds, and dispositions, which require strong interpersonal skills in addition to specialized medical knowledge. To hone these skills, HP education programs typically prepare students through simulation-based training with mannequins or standardized patient interactions with actors, which are effective for teaching hands-on skills and building confidence through realistic practice \cite{kiami_impact_2022}. However, simulation-based learning experiences using mannequins and standardized patients face challenges. They can be resource-intensive and costly, and universities have space constraints and challenges in schedule coordination, limiting the opportunities students have to engage in these learning experiences \cite{so_simulation_2019}.

Virtual Reality (VR) is a promising solution to provide more simulation-based training opportunities in HP education, as its heightened immersion and novel full-body interactivity can create highly engaging training environments tailored to specific needs and used \textit{"whenever and wherever"} \cite{foronda_comparison_2023, pottle_virtual_2019, makransky_cognitive_2021}. Studies show that VR training reduces patient risks, enables frequent practice, and provides contextualized learning experiences \cite{ruthenbeck_virtual_2015}, and that medical VR training simulations are effective across various disciplines, including dentistry \cite{samuel_visuo-haptic_2024, kaluschke_reflecting_2024}, surgery \cite{liu_facilitating_2024, gasques_artemis_2021}, and anatomy \cite{schott_cardiogenesis4d_2023}. Additionally, studies find VR improves students' knowledge acquisition and performance, such as a VR-based childbirth delivery training system that increased learners' knowledge by 24.9\% over traditional mannequin-based methods \cite{liu_facilitating_2024}.

However, previous VR medical simulation training systems have typically utilized rigid procedures and fixed content in order to provide a standardized learning experience \cite{liu_facilitating_2024, muender_evaluating_2022, reinschluessel_versatile_2023}. Such rigidity reduces the potential for repeated use and limits the ability for customization by educators, raising practical concerns for their implementation beyond controlled lab environments \cite{jin_how_2022}. Additionally, previous VR medical simulation training systems often rely on scripted dialogues or step-by-step guidance, limiting their ability to replicate the unpredictability and diversity of real-world medical scenarios \cite{zackoff_impact_2020, samuel_visuo-haptic_2024, sun_design_2023, schott_cardiogenesis4d_2023}. Moreover, past VR medical simulation training systems have not considered the needs of students \cite{tong_exploring_2024}, resulting in systems with low student acceptance \cite{liu_facilitating_2024} and a high learning curve \cite{rebol_evaluating_2023}. 

Thus, this work aims to design a VR medical simulation training system that allows dynamic interaction between HP students and simulated patients utilizing a user-centered design method. As part of a larger study on integrating VR into clinical HP programs, this VR system focuses on interpersonal communication skills in clinical settings, a challenge that clinical faculty described their students face due to limited practice opportunities with authentic patient populations. Research also shows interpersonal communication is a life-long essential skill for HP students, but they face challenges due to the lack of relevant and tailored learning opportunities and poor transfer from training to practice \cite{ammentorp_translating_2022}. 

To design a VR simulation system that addresses these challenges, we conducted semi-structured interviews with advanced HP students who had experience working in a clinical setting through internships or residencies to understand their 1) challenges in communication training and clinical practice and 2) perceptions and expectations in VR simulated communication training. Students described a mismatch between their school-based training and clinical practice, particularly due to wide variation in patient conditions, dispositions, and backgrounds. Given these findings, we are currently co-designing \textit{VAPS}, a \textbf{V}irtual \textbf{A}I \textbf{P}atient \textbf{S}imulator, with clinical professors, which integrates Large Language Models (LLMs) with Embodied Conversational Agents as patients in a hospital setting. Our design leverages the capabilities of LLMs and ECAs to facilitate dynamic interactions and support customizing patients' personalities based on real medical history, meeting learning goals across many HP programs \cite{cassell_embodied_2000}. In this late-breaking work paper, we describe the design of this system to date, along with the insights learned from the user-centered design process. We discuss the implications of integrating LLMs into VR medical simulation training. Given the nature of late-breaking work, we also discuss plans for future iterations of the system and comprehensive evaluation of \textit{VAPS}. 

This work makes a number of contributions to the field of VR in Education 1) We provide valuable design insights from Health Professions students for clinical communication training in VR. 2) We present a novel VR clinical simulation system, \textit{VAPS}, with customizing LLMs-based embodied conversational agents and scenarios. 3) We identify implications from the user-centered design process for future VR+LLMs in medical education.

\section{Interview Study}
As part of a larger project on integrating VR into clinical HP education at a selective US university, we conducted an interview study aiming to answer two research questions: 1) What are HP students' current practices and challenges in clinical communication training? 2) What are their perceptions of using VR for clinical communication training? 

\subsection{Participants}
Our participant criteria required students who had undergone clinical communication training and had at least one instance of on-site clinical communication practice through internships or residencies. Students were recruited by asking HP clinical faculty to refer students and share recruitment materials via email. We conducted interviews with six students (3 Female, 3 Male; M = 25.33 years; SD = 2.21) from three majors: Physician Assistant (PA), Doctor of Pharmacy (PharmD), and Doctor of Physical Therapy (DPT), designated as participants P1 through P6. All participants were familiar with VR in terms of knowing what it is, but there were only two with more extensive use. Participants received a \$10 Amazon gift card for participating. This study was approved by the Institutional Review Board (IRB) of (Blinded) University.

\subsection{Method and Analysis}
Semi-structured interviews consisted of two main parts: (1) experiences in clinical communication training and practice and (2) perspectives on VR for clinical communication training. See Appendix A for the full protocol. Each interview lasted an average of 38 minutes (ranging from 35 to 45). Interviews were conducted in English and were video-audio recorded with participants' consent. Interviews were transcribed and analyzed using the thematic analysis approach \cite{braun2012thematic}. The first author manually reviewed the transcripts for accuracy. Subsequently, two authors independently generated initial codes from all the transcripts using open coding. The authors then discussed their interpretations to reach a consensus and grouped similar codes into clusters. Emerging common themes were identified based on internal connections, and final themes were collaboratively generated using an affinity mapping approach \cite{muller2014Curiosity}.

\subsection{Findings}
\subsubsection{Personalization Challenges during Communication to Diverse Patients}
Our participants identified several challenges during the interviews, all converging on the critical issue of personalization in communication. Students described feeling unprepared for some patient interactions due to the patient's condition, such as pain or visible injuries, emotional state, and diverse backgrounds. For example, one student described difficulty in the early days of their internship because they were not used to seeing patients in severe conditions hooked up to complex machines, as one DPT student said: \textit{``...I think I learned, I got very acclimated to it, and I very much enjoyed it, but when I walked in, [...] it’s my first day, [...] and I saw somebody with end-stage renal disease, and I was there on my first day, that was not something I was prepared for at all...(P5)''} Some other students described challenges interacting with patients who were aggressive or refused to engage with them: \textit{``...Also, in certain settings, the patient could just be like, really overwhelmed, and have like difficulty answering some questions, or there may be like really sensitive topics, that they don't feel comfortable answering, so it can be hard to get information from them in those cases...(P1)''} Our participants also described the diverse backgrounds that patients come from as posing challenges, including treating patients who do not speak English, necessitating an interpreter, or those with low levels of medical literacy, for whom they would need to adapt how they explain conditions and treatments. Overall, students described how patients' conditions, dispositions, and backgrounds varied widely in the contexts they worked in, requiring a great deal of adaptability. Personalizing their communication for all their patient needs was a challenge for them, particularly at the beginning of their clinical work.


\subsubsection{Mismatch between Training and Clinical Communication} 
When asked about the classroom-based training they received for clinical communication, students pointed to limited opportunities to practice authentic communication, and described a mismatch between their training and what they needed to do in practice. Many students described learning on the job and through their clinical experience as the best way to learn the interpersonal communication skills they needed for their profession. While they did receive training on communication in their coursework, they often identified mismatches between that training and what they needed to do in clinical settings. For example, students described simulation lab activities where they interacted with a standardized patient as beneficial practices, but noted it was insufficient because they only used this lab once or twice a year. Furthermore, they described how their practice opportunities were not always realistic in terms of the interactions they would have with patients. For instance, one DPT student described needing to communicate with people unfamiliar with medical terms versus classroom activities where they practiced with their peers who have advanced medical knowledge, as she stated: \textit{``...So [in the lab practical], we're like communicating with our patients. But our patients are classmates. So I feel like just being put in that situation where you have to talk to like an actual patient for that case...(P6)''}. Students also frequently described patients who did not speak English but did not receive training in communicating through interpreters. As one PA student noted: \textit{``...or the patient doesn't really understand what the interpreter saying, or we can't get the exact language that the patient speaks (P1)...''}.

They also described the ways traditional classroom activities were not as helpful due to the evaluation systems. Students described being focused on getting a good grade and following checklists prescribed by their instructors that added pressure to their simulation and role-play activities in classes and interfered with their ability to practice authentically and flexibly, as one PharmD student said:\textit{``...When we do the simulation labs and just like sit in the corner and like... you know that they're (faculty) looking. You know that they're listening to you, and you're trying to just like drown them out, which can make you feel more nervous because you start to think like I need to make sure I'm saying everything to make sure I'm hitting all the points (P4)..."} Grading pressure rendered the training environment less reflective of real clinical settings and, in their view, less meaningful or practical. For example, one PharmD student shared: \textit{``...I think just working with real people, you have a little bit more of a curveball with it. And you just like... everything on that checklist that I used in school, I don't really use that, because it's just different scenarios that are going on...(P3)"} 


\subsubsection{VR as an Opportunity for More Realistic and Varied Practice Opportunities}
During our interviews, most students expressed positive perceptions regarding increased practice opportunities. Notably, participants highlighted the potential of VR as an additional tool for diverse and varied practice scenarios, particularly for extreme situations that are challenging to replicate in standard simulations but are often encountered in real practice. For example, one PA student remarked: \textit{``...It could really be used for like so many different things [...] like, what do I do if a patient starts crying in the middle of a conversation, or how do I bring up bad news or things like that?..(P1)"}. Furthermore, we identified their expectations for various interactive features in VR, including unpredictable conversational dialogues, patient data visualizations, opportunities for self-reflection, and multi-user communication and collaboration. Overall, students identified realistic scenarios and cases to recreate real clinical situations as the most essential factor in VR communication training.

\subsection{Design Insights}
These findings surfaced important insights for designing a VR medical simulation training system. We extracted the following design insights from the representative quotations above: \textbf{(DI1){\textit{Reproduce realistic clinical scenarios:}}} The most important requirement for our VR system is to ensure that the simulation scenarios are as realistic as possible. From the environment to the patients, the system should feature high-fidelity characters, objects, and animations that accurately simulate real-world actions in VR. 
\textbf{(DI2)\textit{Maintain simple interaction to create easy learning experiences:}} 
To help students gain essential skills and confidence, our system should prioritize meaningful VR content over complex or unrealistic interactive features. Since HP students rarely have opportunities to engage with VR, reducing their learning effort through an intuitive, easy-to-use system will enhance their overall training experience.
\textbf{(DI3)\textit{Bring opportunities to allow personalizing simulated scenarios:}} To meet more diverse learning goals, our VR system should have the ability to support customization for a wide range of patient personas and clinical contexts. Thus, users can reuse an identical system that is customized and changed through different learning objectives each time. 
\textbf{(DI4)\textit{Designing communication training as open and unpredictable conversations:}} When simulating communication training in VR, it is essential to go beyond step-by-step, scripted learning approaches. Open and unpredictable conversations reflect the human nature of real interactions, as patients rarely follow a fixed script, therefore making users feel they are engaging with real patients.

\section{VAPS: Virtual AI Patient Simulator}
\subsection{System Overview}
\begin{figure}[htbp]
    \centering
    \includegraphics[width=\linewidth]{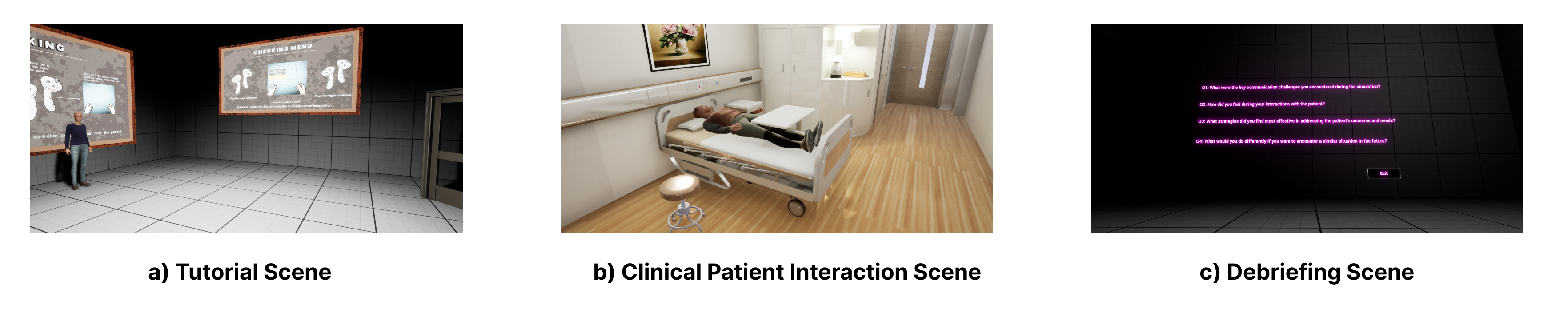}
    \caption{Overview of VAPS: a)Freely-explore tutorial scene; b)A High-fidelity realistic ECA is communicating with users as an AI patient in clinical patient interaction scene; c) Quick self-debriefing science with simple open-questions}
    \label{Overview}
\end{figure}

Driven by the above design insights, we design a novel \textbf{V}irtual reality system empowered by conversational \textbf{A}I \textbf{P}atient \textbf{S}imulator: \textit{VAPS}. The design of this system draws on both the insights identified from the student interview study as well as best practices in HP education simulation, which both emphasize using VR as a practice opportunity to apply prior knowledge, not a didactic instructional system. The system therefore has three scenes to simulate the entire process: tutorial, clinical patient interaction, and reflection. The goal of this system is to enhance HP students' clinical communication skills by providing realistic and dynamic practice opportunities. 
In the \textbf{\textit{tutorial}} scene, users are introduced to interacting with the VR system. They are guided through interactive guidelines to explore and familiarize themselves with the interaction of \textit{VAPS} with the help of a high-fidelity ECA "tour guide." In this scene, the learner is also pre-briefed on the patient case and learning objectives. In the \textbf{\textit{clinical patient interaction}} scene, we designed another ECA to act as a patient lying in a simulated hospital room (Fig.\ref{Overview}). Learners enter the room and must sit near the patient's bed to collect their history and details about their current condition. After this interaction, learners must identify a referral or consultation for the patient, then enter the final \textbf{\textit{reflection}} scene. In this scene, students reflect on their performance based on standardized prompts and align with simulation education standards.
\subsection{Embodied Conversational Agents}
Inspired by our \textit{DI1, DI3} and \textit{DI4}, we determined the capabilities of Large Language Models (LLMs) would be the optimal solution to address the need for dynamic and customizable conversations, and chose ECAs as the vehicle to deploy LLMs. This aligns with previous research findings, which suggest that combining non-verbal conversational and empathetic cues—such as hand gestures and facial expressions—with spoken dialogue can significantly enhance user trust in ECAs \cite{bickmore_relational_2001}. Thus, our ECAs have the following features informed by design insights, \textbf{\textit{1) Personalized ECAs Design}}: Our ECAs can be trained from prompts in the patient's medical history and personal backstory. Also, users can choose their personality traits, state of mind, preferred language, and speaking style during the design process. \textbf{\textit{2) Integration with Medical Records:}} Our ECAs support uploading real medical files to a knowledge repository, ensuring accurate and context-sensitive responses. \textbf{\textit{3) Dynamic Narrative Design:}} Our ECAs can be designed with narrative flow. Users could design ECAs to adapt to any triggers or decisions during interactions. \textbf{\textit{4) Realistic Animations:}} Our ECAs perform basic head movement and hand gestures, as well as synchronized lip motions, to create human-like, realistic interactions during communication.

The design of the clinical patient interaction scenario draws on best practices in simulation for HP education, and we are co-designing this with a clinical professor with over 15 years of experience designing clinical simulation training. While the prototype begins with one scenario and a limited number of patient personas for learners to practice (see Section 3.4), ultimately, the system will allow clinical faculty to customize the simulation based on their course objectives and learner needs. To address the challenges faced by non-AI experts, such as clinical professors, in creating prompts for ECAs' every feature, we developed a user-friendly design form (See  Appendix B). For example, clinical professors can input patient backstories that are common materials for simulation exercises and select levels for the patient's personality in the design form, which is then converted into a patient within the \textit{VAPS} system by the development team. This also emphasizes the importance of reviewing scripts manually to ensure the appropriateness of training prompts. Finally, the LLMs trained for \textit{VAPS} incorporate patient profile information into profile-specific embodied character models.


\subsection{System Interaction}
Our interaction design, inspired by \textit{DI2}, emphasizes simplicity for ease of learning. Thus, users can engage with the system in \textit{VAPS} through three primary actions: Move Joystick (to teleport to a specific location), Push Button (to talk, inspect, or reset), and Press Trigger (to interact with buttons in the VR environment). Each button has a distinct purpose, enabling versatile interactions. For example, the right controller's `A' button enables voice communication, while the `B' button shows a patient information board with key details like medical history and demographics. These straightforward interactions are designed to minimize learning barriers and enhance usability. 

\subsection{Example Learning Case}
The initial prototype utilizes one learning case that illustrates how \textit{VAPS} can be integrated into HP education. In this scenario, the learner must conduct a subjective interview with a patient who is diagnosed with chronic obstructive pulmonary disease (COPD), a progressive lung disease. This diagnosis is a good candidate for the prototype because professionals from all the target clinical programs would need to treat a patient with this disease, namely nurses, physician's assistants, physical therapists, and pharmacists. Our ECAs allow learners from these varied professions to conduct the relevant patient interview, which may be focused more on physical activity or medications. The patient's diagnosis and condition remain constant for the simulation, but the patient persona will vary according to four common challenges students face: 1) a highly emotional patient, 2) an aggressive or rude patient, 3) a patient who is refusing care, and 4) a patient with low medical literacy. Learners will therefore be able to practice conducting the subjective interview multiple times to adapt to different patient conditions. The use of LLMs allows the interaction to vary dynamically in each use, and the reflective activity helps them identify areas of improvement between each round of practice.

The proposed implementation would allow students to practice with \textit{VAPS} on their own, for example, homework, and then have a reflective activity with their instructor or a small group of peers. This allows for more in-depth debriefing, aligning with standard practices in simulation, as typically conducted following an in-person simulation session. This learning case serves as just an illustrative example rather than a prescriptive guide for implementing \textit{VAPS} in actual learning contexts, but ultimately, \textit{VAPS} will facilitate repeated use by putting design in the hands of instructors. Clinical professors can modify the scenario and the patient personas to introduce different learning objectives aligned with their courses through the design form. We discuss more learning scenarios and the vision of \textit{VAPS} in Section 4.2.

\subsection{Implementation}
The prototype system was developed by the research team using Meta Quest Pro and Meta Quest 3, though it is compatible with any immersive VR headset. The system leverages Unreal Engine 5, chosen for its versatility and suitability for immersive VR applications. The ECAs were created using MetaHuman\footnote{https://www.unrealengine.com/en-US/metahuman} as a high-fidelity model and character detail. 
We used Convai Api\footnote{https://convai.com/} to enable communication, customization, animation, and LLMs integration (GPT-4o, generally regarded as the state-of-the-art LLM) employed in \textit{VAPS}.
The 3D hospital room scene was constructed using online open-source resources. For the reflection and tutorial rooms, we utilized the standard Unreal VR template scenarios, prioritizing the content within these steps over the environmental details.

\section{Discussion}
\subsection{Supporting Customized VR Learning with LLMs}
Our interview findings highlight challenges health professionals face when communicating with highly dynamic patients in clinical settings, which students reported they need more practice on before entering clinical practice. This informs a key design implication for VR simulation systems: the need to support customizable scenarios that address diverse learning objectives. Previous literature has highlighted the potential of LLMs in educational and social VR environments \cite{liu_classmeta_2024, wan2024building} as well as the advancement of ECAs in clinical VR systems \cite{chheang_towards_2024, laranjo2018conversational}. Building on these insights, we combine these two elements—LLMs and ECAs—into the design of a VR simulation to support communication skills practice in a more dynamic clinical setting. Our co-design process also led to the identification of LLMs as a powerful tool to address a felt need in health professions education of customization need, rather than simply following the current \textit{"LLMs + Everything"} trend. \textit{VAPS} employs LLM-based ECAs as direct objects of human-AI interaction to facilitate learning rather than mediating interactions between humans or non-intelligent agents. Compared to VR, other mobile applications such as computer screens or iPads, may offer greater accessibility in clinical training and require less technical expertise to interact with LLMs. However, given the use of ECAs as vehicles and the focus on communication skills as a learning goal in this paper, we utilize VR because these aspects can be simulated more naturally in an immersive environment. Future research could compare the specific performance of different vehicles in this context.

An important feature of \textit{VAPS} is its flexibility. While the case of communicating with a patient diagnosed with COPD described in Section 3.4 will be our initial test, the system allows for instructors to determine other patient conditions and upload their medical history. The personas of the patients are also customizable to provide students with varied practice with realistic patients. This flexibility makes VAPS stand out from previous VR medical simulation systems, as it allows students to encounter the unpredictability and diversity of real-world medical scenarios while enabling faculty to customize different learning goals without rigid instructional constraints.
Notably, we recognize that AI ethics may raise potential concerns regarding data privacy. Therefore, we currently do not allow instructors or students to directly upload data into the system, instead, all real data undergoes a secondary manual review to ensure it adheres to standardized simulation annotations. Based on our experience, we strongly advocate for researchers to place greater emphasis on censorship and protection to ensure data privacy and mitigate LLM hallucination issues. A better solution is needed in this rapidly evolving domain. Further, the implementation of the tool is adaptive to instructors' needs. \textit{VAPS} could be used before or after in-person simulation experiences to provide additional practice opportunities. It may be used as homework by individual students who then debrief their performance in class, or instructors may ask small groups of students to use it in class. In the future, we aim to add more features to \textit{VAPS} that will broaden its applications, including outputs like video and text summaries that can enable reflection and instructor feedback, multiplayer modes where students can practice interprofessional communication that include other healthcare professionals, and options for communicating via interpreters when treating patients who are not fluent in English. With these capabilities, \textit{VAPS} will enhance customization meeting various needs in clinical communication training.

\subsection{Leveraging VR to Expand Realistic and Dynamic Training Opportunities}
Our interview findings align with \citet{yarmand_id_2024}, indicating a mismatch between training and clinical practice. Merely replicating traditional simulation practices in VR risks reproducing this mismatch. However, our findings illustrate how students perceive VR as a promising medium for providing realistic and dynamic practice opportunities. Therefore, \textit{VAPS} focuses on dynamic training procedures that mimic unpredictable situations encountered in real clinical practice, enhanced by high-fidelity animation, textures, and models. Due to most students and faculty being unfamiliar with VR, \textit{VAPS} prioritizes simple interactions and environments to avoid distracting learners or making them inaccessible. 

During the development of our system and the training of ECAs, we encountered both the challenges and importance of persona development, as suggested by \citet{chin_like_2024}. \textit{VAPS} addresses these challenges by helping clinical professors design prompts, something that non-AI experts typically struggle with  \cite{Yang2024Johnny}. Doing so required collaboration with clinical faculty to design the ECAs and the development of a form that simplifies the LLMs training process. Our system allows faculty to contribute domain-specific information and features that increase realism including narrative flow, medical records, and personality metrics. This promotes dynamic practice that is aligned with the curriculum and addresses limitations in traditional VR clinical simulations, which often rely on a rigid and fixed learning structure \cite{liu_facilitating_2024, muender_evaluating_2022, reinschluessel_versatile_2023}. However, we acknowledge our system has some limitations. These include a focus primarily on communication training and a fixed VR environment, which restricts the ability to simulate diverse clinical scenarios, such as a physical therapist treating patients in rural hospitals. Thus, we advocate that future work should not only build upon this foundation to support more clinical scenarios but also support interprofessional education, such as enabling multiple HP students from various disciplines to study together.
Additionally, with advancements in generative AI for 3D modeling \cite{Lei2024VRCopilot} and human animation \cite{Azadi_2023_ICCV}, we are optimistic that these constraints can be addressed in the future. However, bringing VR to higher educational institutions ultimately requires efforts from the entire network of stakeholders \cite{jin_how_2022}.

\section{Conclusion and Future Work}
In this paper, we explored the design of LLM-based ECAs for VR clinical communication training. We began with an interview study involving six Health Profession students to examine current clinical communication training practices and their perceptions of VR training. Based on the findings, we proposed a set of design insights and developed a prototype VR system, \textit{VAPS}, to address the identified design requirements. Furthermore, we highlighted the significance of customization and realism in VR simulation systems. For future work, we plan to evaluate \textit{VAPS} in the real learning context. Our evaluation roadmap includes conducting an initial usability test followed by long-term studies to assess the effectiveness and performance of VR-based simulation training in comparison to traditional simulation training, specifically in the domain of clinical communication.


\begin{acks}
We are grateful for all the students and faculty who participated in this research. We are also grateful to Ara Jung from the ReGameXR Lab for her assistance. This research was supported by a Northeastern University TIER 1 interdisciplinary research grant.
\end{acks}

\bibliographystyle{ACM-Reference-Format}
\bibliography{main}

\appendix
\section{Interview Protocol}

\paragraph{Part One: Interviewee Background (5-10 minutes)}

\begin{itemize}
    \item Can you tell me what program you are in? What year are you in that program?
    \item Why did you choose to enroll in this program in college?
    \item What health professions courses have you taken at (Blinded) University?
    \item What are your experiences and knowledge about Virtual Reality (VR)?
    \item What clinical training and practice did you receive?
\end{itemize}

\paragraph{Part Two Interview Questions (30-40 minutes)}

\begin{itemize}
    \item Tell me about any training you have received in communication in clinical settings, like delivering information to patients or having conversations with them. What did you learn? How was the training conducted?
    \subitem a.	During the training, what challenges did you face with the current practices, and why?
    \subitem b.	Is there anything in your training process that you think can be improved?
    \item In your work and studies, do you need to collaborate with other clinicians from different disciplines? Have you received any training about that?
    \subitem a.	If yes, what challenges have you encountered when you collaborate with other professions?
    \subitem b.	How do you balance communicating with patients and other clinicians?
    \subitem c.	If no, can you imagine what happens when you would need to work with a clinician from another discipline?
    \item What do you think is the most helpful method for learning how to communicate in clinical settings?
    \subitem a.	During your communication with patients, which aspects would you value more?
    \item What are your thoughts on how virtual reality could be integrated into your education for training on communication in clinical settings?
    \subitem a.	If positive, what expectations do you think virtual reality can help achieve that traditional methods cannot in your learning? And what interactive features you would like to use?
    \subitem b.	If not, why do you think this is not feasible and what kind of effort is needed?
    \item Is there anything else that you would like to contribute that we have not discussed?
\end{itemize}

\section{Design Forms}
Fill out this form to design an AI patient in VSPA!

\textbf{
Medical History:}
\begin{itemize}
    \item Reason for Visit:
    \item Past Medical History:
    \item Family Medical History:
    \item Current Medications:
    \item Allergies:
    \item Smoking:
    \item Alcohol Use:
    \item Exercise Habits:
    \item Diet:
\end{itemize}

\textbf{Information:}
\begin{itemize}
    \item Name:
    \item Age:
    \item Gender Identity:
    \item Ethnicity/Race:
    \item Language(s) Spoken:
    \item Accent:
    \item Occupation:
\end{itemize}

\textbf{Personal backstory:}

\textit{Describe the character's backstory as instructions in second person i.e. phrases like "You are.., You are an expert on.., Your life is based on the following facts…" Try to keep the backstory limited to a short background of the character and the world / scenario they are present in, basic personality and mannerisms. For longer factual information, use the knowledge bank.}

\begin{figure}[htbp]
    \centering
    \includegraphics[width=\linewidth]{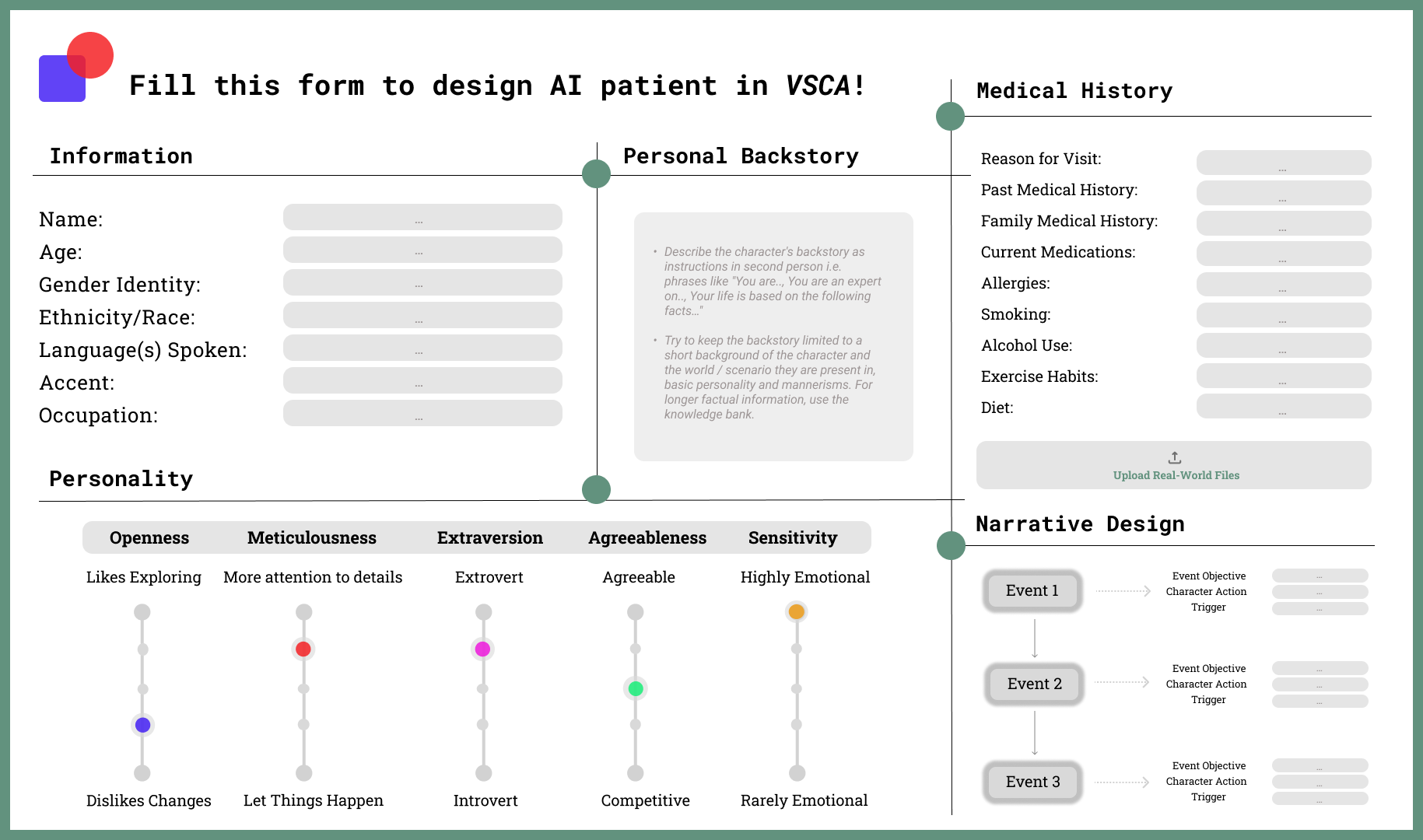}
    \caption{A user-friendly design form for helping non-AI experts design the features and characters of AI patients in \textit{VAPS}}
    \label{design form}
\end{figure}

\textbf{Personality}

score with 1-5 for each characteristic, see Fig \ref{design form} 

\textbf{Narrative Design (Optional)
}
\begin{itemize}
    \item Event l
    \subitem Event Objective
    \subitem Character Action
    \subitem Trigger
    \item Event 2
    \subitem Event Objective
    \subitem Character Action
    \subitem Trigger
    \item Event 3
    \subitem Event Objective
    \subitem Character Action
    \subitem Trigger
\end{itemize}

\section{Example Prompts of Training ECAs by Clinical Professor}
\textbf{Reason for Visit:} 
Admitted to the hospital 2 days ago following a severe COPD exacerbation triggered by a respiratory infection. Symptoms included severe shortness of breath, wheezing, productive cough, and low oxygen saturation (SpO2 85\% in air in the room).

\textbf{Past Medical History: }
\begin{itemize}
    \item Chronic obstructive pulmonary disease (COPD) – diagnosed 7 years ago.
    \item Hypertension – controlled with medication.
    \item Type 2 diabetes mellitus – managed with oral agents.
    \item Anxiety – mild, occasional symptoms managed without medication.
    \item History of recurrent bronchitis.
\end{itemize}
Family Medical History:
    \begin{itemize}
        \item Father: Passed away at 68 from lung cancer (smoker).
        \item Mother: History of type 2 diabetes and hypertension.
        \item Brother: Diagnosed with COPD, current smoker.
    \end{itemize}
Current Medications:
    \begin{itemize}
        \item Albuterol inhaler (as needed).
        \item Fluticasone/salmeterol (Advair Diskus) – twice daily.
        \item Tiotropium (Spiriva) – once daily.
        \item Metformin – 1,000 mg twice daily.
        \item Lisinopril – 10 mg daily.
    \end{itemize}
Allergies:
    \begin{itemize}
        \item Sulfa drugs (causes rash and swelling).
    \end{itemize}
Smoking:
    \begin{itemize}
        \item 50 pack-years (smoked 1 pack per day for 50 years). Quit 2 years ago after his first hospitalization for COPD exacerbation.
    \end{itemize}
Alcohol Use:
    \begin{itemize}
        \item Occasionally drinks beer on weekends, 1-2 bottles per occasion.
    \end{itemize}
Exercise Habits:
    \begin{itemize}
        \item Minimal activity due to breathlessness but enjoys short walks around his neighborhood when able.
    \end{itemize}

Diet:
    \begin{itemize}
        \item High in carbohydrates (e.g., rice, beans, tortillas). Loves fried foods and sugary drinks. Attempts to follow dietary recommendations for diabetes but struggles with portion control.
    \end{itemize}

\textbf{Information:}
\begin{itemize}
    \item Name: Samuel "Sam" Rivera
    \item Age: 62
    \item Gender Identity: Male (he/him/his)
    \item Ethnicity/Race: Hispanic/Latino
    \item Language(s) Spoken: Primarily Spanish; conversational English.
    \item Accent: Noticeable Spanish accent with a deliberate and slow speech pattern in English.
    \item Occupation: Retired auto mechanic.
\end{itemize}

\textbf{Personal backstory:}

Sam is a retired auto mechanic who worked in the same shop for over 35 years. He takes great pride in his work and often reminisces about the cars he fixed and the friends he made. He lives in a small apartment with his dog, Luna, who is his primary companion. His wife passed away 10 years ago from breast cancer, and he has two adult children who live out of state.
Sam identifies as a devout Catholic and often prays for strength when facing challenges. He struggles with feelings of loneliness, particularly since his children visit infrequently. Despite this, he maintains a positive outlook most of the time and is well-liked by his neighbors, who occasionally check in on him. He speaks Spanish at home and prefers to communicate in Spanish, although he can manage in English for basic conversations. He expresses frustration when medical instructions are not clear or when he feels rushed during interactions. Sam is cooperative with medical staff but admits to feeling overwhelmed by his condition and the numerous medications he needs to take. He often relies on his daughter, who lives in Florida, for emotional support over the phone.

\textbf{Personality}
\begin{itemize}
    \item Openness 2
    \item Meticulous 3
    \item Extraversion 2
    \item Agreeableness 3
    \item Sensitive 3
\end{itemize}

\textbf{Narrative Design}

Event l Event Objective
\begin{itemize}
    \item Demonstrate empathy and active listening.
    \item Provide appropriate emotional support.
    \item Collaborate with other team members to address mental health needs.
    \item Character Action: Crying and Upset Patient
\end{itemize}
    Trigger: 
    
    During session, Sam becomes tearful and expresses feelings of hopelessness about his condition, saying, "I can't do this anymore. I'm a burden to everyone."
Key Points for Each Discipline:
\begin{itemize}
    \item Physical Therapy: Pause the session, provide reassurance, and adjust the plan to match his emotional and physical capacity.
    \item Nursing: Explore underlying emotional concerns, offer resources for mental health support, and coordinate with the care team.
    \item Pharmacy: Discuss the potential need for anxiolytic or antidepressant therapy.
    \item SLP: Assess if emotional distress impacts swallowing safety and ability to participate in therapy.
    \item PA: Validate his concerns, provide support, and potentially consult psychiatry or counseling services.
\end{itemize}

Event 2
Event Objective
    \begin{itemize}
        \item De-escalate the situation calmly and professionally.
        \item Maintain boundaries while addressing the patient’s concerns.
        \item Recognize potential triggers and underlying issues for anger.
        \item Character Action Angry and Inappropriate Behavior

    \end{itemize}
Trigger: 

Sam becomes agitated during a nursing assessment, yelling, "You people don’t know what you’re doing! This place is useless!" He makes a derogatory comment toward a staff member.
Key Points for Each Discipline:
    \begin{itemize}
        \item Nursing: Use a calm tone, acknowledge his frustration, and attempt to identify specific complaints.
        \item Physical Therapy: Adjust the session as needed, ensuring safety and addressing barriers to cooperation.
        \item Pharmacy: Assess if medications could contribute to irritability (e.g., corticosteroids).
        \item SLP: Evaluate whether difficulty communicating or swallowing might be contributing to his frustration.
        \item PA: Lead a debrief with the care team to align approaches and discuss possible psychiatric or behavioral interventions.
    \end{itemize}

Event 3
Event Objective 
    \begin{itemize}
        \item Communicate effectively with a patient who is reluctant or overwhelmed.
        \item Assess and adapt health education for low medical literacy.
        \item Encourage engagement in care while respecting autonomy.
        \item Character Action Refusing Care and Low Medical Literacy
    \end{itemize}
Trigger: 

Sam refuses to attend therapy sessions and states, "What's the point? My lungs are shot anyway." He also struggles to understand instructions for his inhaler, stating, "I don't need all these fancy gadgets."
Key Points for Each Discipline:
    \begin{itemize}
        \item Physical Therapy: Use motivational interviewing techniques to explore barriers and encourage participation.
        \item Nursing: Simplify health education materials and provide hands-on demonstrations for self-management.
        \item Pharmacy: Explain medication benefits in simple, relatable terms, and demonstrate inhaler techniques.
        \item SLP: Ensure that instructions are clear and accessible, considering potential swallowing or cognitive barriers.
        \item PA: Address his refusal empathetically, emphasizing achievable goals and the importance of adherence to the care plan.
    \end{itemize}

\end{document}